\documentclass[preprint2]{aastex}

\begin{document}

\title{The Proper Motion of the Neutron Star in Cassiopeia A}


\author{Tracey DeLaney and Joseph Satterfield}
\affil{Physics \& Engineering Department, West Virginia Wesleyan College, 
Buckhannon, WV 26201, USA}
\email{delaney\_t@wvwc.edu}

\begin{abstract}

Images from the High Resolution Camera on the \emph{Chandra X-ray Observatory} 
were used to measure the proper motion of the neutron star in Cassiopeia A 
over a baseline of 10 years (1999-2009).  One background source and 13 
quasi-stationary flocculi were chosen to register the two images.  Pixel 
offsets between features at each epoch were measured using four different 
statistical methods: Gaussian fitting, centroiding, cross correlation, and the 
Cash statistic.  In many cases the offset measurements disagree in magnitude 
and/or direction by as much as 1\arcsec, resulting in large uncertainties.  As 
a result, the measurement for the motion of the neutron star is marginal at 
390$\pm$400 km s$^{-1}$ in the southeast direction. This motion is typical of 
the birth velocities of young pulsars and is consistent with the inferred 
proper motion based on the offset of the neutron star from the center of 
expansion of the optical ejecta.  

\end{abstract}



\section{Introduction}

\subsection{General Background on Cas A}

Cassiopeia~A (Cas~A; 3C 461, G111.7-2.1) is the 2nd-youngest-known supernova 
remnant (SNR) in the Galaxy and lies at a distance of 3.4 kpc away \citep{rhf95}.  
With the discovery of light echoes from the explosion, we now know that Cas~A resulted 
from an asymmetric type IIb explosion \citep{kbu08,rfs11}.  One of the most exciting 
discoveries with the \emph{Chandra X-ray Observatory} was the neutron star in Cas~A 
\citep{tan99}.  The neutron star does not show X-ray or radio pulsations 
\citep{mcl01,mer02} and lacks a synchrotron nebula \citep{hwa04} typical of ordinary 
young pulsars still located within their supernova remnants (SNRs).  \citet{hh09} have 
shown that the neutron star is successfully fitted using a low magnetic field carbon 
atmosphere model and emits thermal radiation from the entire surface.  Furthermore, the 
cooling rate of the neutron star has been measured \citep{hh10}, it has a gravitational 
mass of M$\approx$1.3-2 M$_{\sun}$ \citep{yhs11}, and the neutrons in the core have 
recently become superfluid \citep{syh11}. 

The neutron star is offset from the well determined expansion center of the 
undecelerated optical ejecta by $7\arcsec$ nearly due south \citep{fes06b}.  
The implied projected velocity based on this offset is about 350 km s$^{-1}$ (assuming 
an age of 330 years) which is typical of young pulsars \citep{lyn94}.  However, the 
offset is $\approx90\degr$ from the axis defined by the northeast and southwest jets 
\citep{hwa04} -- the axis along which one might expect a ``birth kick'' due 
to supernova explosion asymmetries \citep{lai01}.

The jets, however, do not provide the only symmetry axis in Cas~A.  The X-ray iron 
emission and the infrared neon emission also fall into bipolar structures to the north 
and south/southeast \citep{err06,srd09,drs10}.  Aligned with the north/south neon axis 
are gaps in the distribution of outer optical ejecta knots \citep{fes06b}, further 
supporting a roughly north-south symmetry axis.  These various bipolar symmetries 
observed in multiple wavebands in the north-south direction provide an alternative 
predicted kick direction for the neutron star -- one that agrees with the inferred motion.

Although the offset of the neutron star from the expansion center of Cas~A is effectively 
irrefutable evidence of the direction and magnitude of motion, a simple verification of 
the motion of the neutron star is in order.  Therefore, in this paper we report on the 
proper motion measurement of the neutron star using images from the High Resolution 
Camera (HRC) on the \emph{Chandra X-ray Observatory}.  In \S\ref{obs} we describe the 
observations.  In \S\ref{register} we discuss the image registration between epochs with 
the calculation of the neutron star motion presented in \S\ref{motion}.  In \S\ref{disc} 
we discuss the results from our analysis and we offer concluding remarks in \S\ref{conc}.

\section{Observations}
\label{obs}

\begin{deluxetable}{rrrrrr}
\tablewidth{0pt}
\tablecaption{Observational Parameters \label{obsparms}}
\tablehead{
\colhead{} & \colhead{} & \colhead{RA} & \colhead{Dec} & \colhead{Roll} & \colhead{Livetime}\\
\colhead{Obs Id} & \colhead{Date} & \colhead{(deg)} & \colhead{(deg)} & \colhead{(deg)} & \colhead{(sec)}}
\startdata
1505 & 1999 Dec 19 & 350.8566 & 58.81010 & 287.1300 & 48720 \\
11240 & 2009 Dec 20 & 350.8568 & 58.81059 & 287.0352 & 12914 \\
12057 & 2009 Dec 13 & 350.8570 & 58.81053 & 287.0350 & 10884 \\
12058 & 2009 Dec 16 & 350.8571 & 58.81051 & 287.0349 & 9224 \\
12059 & 2009 Dec 15 & 350.8568 & 58.81058 & 287.0351 & 12801 \\
\enddata
\end{deluxetable}

New observations of Cas A were taken in Dec 2009 with the HRC instrument
on \emph{Chandra}.  Care was taken to use the same pointing and roll parameters 
as the original 50-ks HRC observation in Dec 1999 in order to mitigate 
any effects from the asymmetrical point spread function.  As shown in 
Table~\ref{obsparms}, the 2009 observations were taken over four separate days and 
totaled about 3 ks less than the 1999 observation.  The pointing centers for the 
observations varied by only a few arcseconds and the roll only changed by 0.1\degr.  
Before merging the four level two event files from the 2009 observations, Gaussian 
fitting of the neutron star was used to determine that the registration uncertainty 
was less than 0.1 pixel, where 1 HRC pixel=0\farcs1318.  The 1999 data were re-processed 
using the same version of Ciao and CALDB as the 2009 data, including aspect correction.  

\section{Image Registration}
\label{register}

\subsection{Registration Sources}

\begin{figure}[t]
\epsscale{1.0}
\plotone{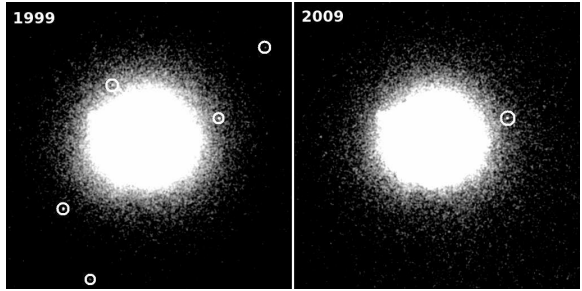}
\caption{Background point sources in the field of view of the HRC in 1999 
(left) and 2009 (right).  We have named the lone background source in the 2009 image 
src2.  
\label{extrefs}}
\end{figure}

The original intent was to use five background sources within the field of view of the 
1999 HRC image to register to the 2009 image.  Four of the five sources were present on 
the ROSAT image of Cas~A, leading us to believe that they were robust sources to use.  
Unfortunately, as Figure~\ref{extrefs} shows, only one of these background sources was 
present in the 2009 HRC image.  We have named this lone background source src2.

Therefore, we needed to use Cas~A to register the 1999 and 2009 HRC images.  The proper 
motions of the X-ray ejecta in Cas~A are on the order of a few thousand kilometers per 
second \citep{drf04}.  The X-ray proper motion measurements were made over a span of only 
two years and used the neutron star as the primary registration source.  Therefore, there 
may be asymmetries or measurement artifacts in the proper motions that could result in 
errors of order a few hundred kilometers per second, which is on the order of the 
inferred velocity of the neutron star. 

\begin{figure}[b!]
\epsscale{1.0}
\plotone{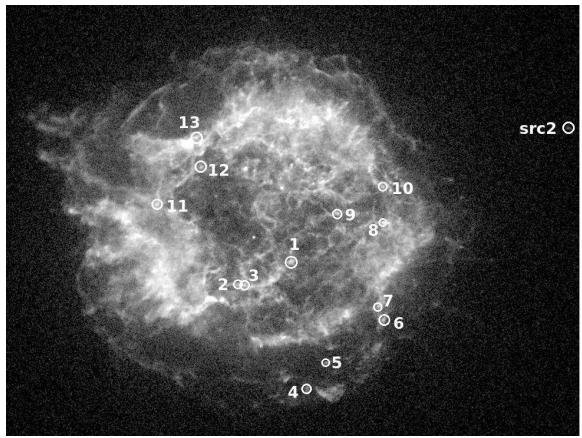}
\caption{The QSFs and background point source used for image registration. \label{regions}}
\end{figure}

We then decided to use a number of quasi-stationary flocculi (QSFs).  QSFs are 
slow-moving ($v\lesssim$ 500 km s$^{-1}$) stellar wind clumps with well-determined 
velocities based on over 30 years of optical observations \citep{kb76,bk85}.  
\citet{drf04} found that many of the optical QSFs had direct X-ray counterparts and 
that X-ray QSFs could be identified based on their spectra and dynamics.  Thirteen QSFs 
were chosen based on their brightness, isolation in the X-ray image, availability of 
optical and/or X-ray proper motion information, and position on the image -- we wanted 
to have about equal numbers of QSFs above/below and left/right of the neutron star.  
These QSFs are identified in Figure~\ref{regions}.  

\subsection{Measuring Offsets between Epochs}

There are many statistical techniques for finding the ``center'' of an emission clump or 
determining the offset between two clumps.  All of these techniques require the setting 
of some sort of bounding box over which the statistical calculations are performed.  In 
addition, there are a range of binning and smoothing parameters that can be applied 
to the images before they are analyzed.

To determine how much of an effect the bounding box, binning, and smoothing had on our 
analysis, we used the Gaussian fitting routines IMFIT and JMFIT within the Astronomical 
Image Processing System (AIPS) with a range of these parameters.  IMFIT and JMFIT use 
2-dimensional elliptical Gaussians and are able to solve for a constant, linear, or 
quadratic two-dimensional baseline surface.  Images of each region at each epoch were 
made with a range of binning from none to a factor of 4 and with a range of Gaussian 
smoothing sizes from no smoothing to smoothing over 2\farcs1.  Bounding boxes 
ranged in size from twice the size of the clump to hugging the clump tightly and a range 
of parameters were chosen to describe the baseline surface.  The two AIPS Gaussian 
fitting routines were very robust in reporting clump centers with the poorest standard 
deviation of about 0.5 HRC pixel (0\farcs0659) for QSF4 and src2.

We repeated the analysis above using the centroiding feature within SAOImage DS9.  
Centroiding was quite robust, with similar standard deviations to the Gaussian fitting. 
However, in many cases the clump centers were significantly different from those 
determined based on Gaussian fitting.  This was to be expected because centroiding and 
Gaussian fitting weight image data quite differently.

Our final two statistical measures computed the registration offsets between each epoch 
directly.  For the first technique, the 1999 image of each region was shifted in $x$ and 
$y$ with respect to the 2009 image and the Cash statistic \citep{cash} was computed 
between the two images at each shift position.  The amount of the shifting 
and the size of the box over which the Cash statistic was computed varied from 20-80 HRC 
pixels (2\farcs6-10\farcs5) depending on the size of the region and nearby confusing 
emission structures.  Finally, the AIPS routine CONVL was used to determine the offset 
between images at each epoch using cross correlation.  We again found that the solutions 
were highly dependent on the statistical method used with much less dependence on 
binning, smoothing, and bounding box.  

\begin{figure}[t]
\epsscale{1.0}
\plotone{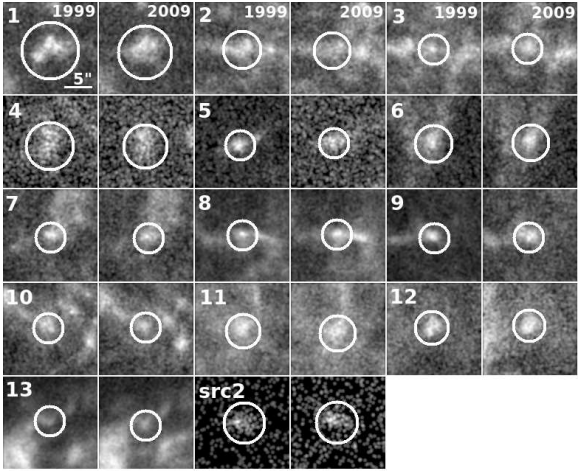}
\caption{1999 to 2009 comparison of regions used for image registration. No binning has 
been applied, but the images have been smoothed to 0\farcs4. \label{regcomps}}
\end{figure}

The single most complicating factor in these position and offset measurements is that the 
QSFs and src2 changed morphology between 1999 and 2009, as shown in 
Figure~\ref{regcomps}.  As a result, the different statistical measures varied not only 
in magnitude of offset between 1999 and 2009, but in direction of offset as well with the 
largest difference being about 8 HRC pixels (1\farcs1).  Columns 5-12 of 
Table~\ref{regdata} show the results of the individual statistical techniques.  The 
offsets reported for each region are averaged over all of the binning, smoothing, and 
bounding box parameters applied as discussed above and are reported in HRC pixels.  The 
$x$- and $y$-offsets indicate how far the 1999 region is shifted from the 2009 region 
with negative numbers indicating offsets to the east or south.

We also report in Table~\ref{regdata} the X-ray counts for each region as a measure of 
brightness.  In principle, one would expect low brightness regions to have a larger 
uncertainty than high brightness regions.  However, the morphology changes so dominated 
the offset measurements that there was no correlation between counts and measurement 
scatter.

\begin{deluxetable}{rrrrrrrrrrrrrrrr}
\rotate
\tablecolumns{16}
\tabletypesize{\scriptsize}
\tablewidth{0pt}
\tablecaption{Data used to register images. \label{regdata}}
\tablehead{
\colhead{} & \colhead{} & \multicolumn{2}{c}{Proper Motion\tablenotemark{a}} & \multicolumn{2}{c}{Cash Statistic} & \multicolumn{2}{c}{Gaussian} & \multicolumn{2}{c}{Centroid} & \multicolumn{2}{c}{Cross Correlation} & \multicolumn{2}{c}{Mean\tablenotemark{b}} & \multicolumn{2}{c}{Uncertainty\tablenotemark{c}} \\
\colhead{Region} & \colhead{Counts} & \colhead{$x$} & \colhead{$y$} & \colhead{$x$-offset} & \colhead{$y$-offset} & \colhead{$x$-offset} & \colhead{$y$-offset} & \colhead{$x$-offset} & \colhead{$y$-offset} & \colhead{$x$-offset} & \colhead{$y$-offset} & \colhead{$x$-offset} & \colhead{$y$-offset} & \colhead{$x$} & \colhead{$y$}} 
\startdata
src2 & 2039 & \nodata & \nodata & 0.00 & -4.00 & 2.20 & 3.90 & 1.24 & 1.90 & 2.00 & -2.92 & 1.36 & -0.28 & 0.50 & 1.89 \\
1 & 8561 & \emph{-7.59} & \emph{-0.76} & -2.75 & -4.75 & -4.65 & -3.32 & 0.47 & 0.23 & 0.26 & -0.15 & 5.92 & -1.24 & 1.24 & 1.21 \\
2 & 8336 & -0.27 & 1.18 & 2.00 & 1.50 & -2.65 & -1.12 & -0.24 & -0.91 & 0.10 & 0.88 & 0.07 & -1.09 & 0.95 & 0.65 \\
3 & 2771 & -0.27 & 1.18 & 0.75 & -0.25 & -2.29 & -0.13 & -0.24 & -0.68 & 0.61 & 0.24 & 0.02 & -1.39 & 0.70 & 0.19 \\
4 & 908 & \emph{-1.52} & \emph{-0.76} & 1.25 & 0.75 & -2.62 & -0.06 & 5.92 & 0.61 & -0.92 & -6.00 & 2.43 & -0.42 & 1.85 & 1.62 \\
5 & 842 & \emph{3.79} & \emph{-0.76} & -2.00 & 0.00 & -3.15 & -0.67 & -4.14 & -0.38 & 2.52 & 0.95 & -5.48 & 0.74 & 1.47 & 0.35 \\
6 & 1682 & \emph{2.28} & \emph{1.52} & 0.00 & 0.75 & -2.43 & -1.53 & 0.30 & 0.15 & 1.06 & -0.38 & -2.55 & -1.77 & 0.75 & 0.48 \\
7 & 3044 & -0.73 & 1.67 & 2.25 & 1.75 & 0.87 & 2.33 & -4.14 & -0.76 & 1.02 & 0.73 & 0.73 & -0.66 & 1.41 & 0.68 \\
8 & 3103 & \emph{1.52} & \emph{1.52} & 1.00 & 0.75 & 2.04 & -3.25 & 0.80 & 0.38 & 1.55 & -0.08 & -0.15 & -2.07 & 0.27 & 0.92 \\
9 & 3560 & -1.75 & 0.90 & -0.50 & 1.00 & -0.77 & -5.64 & -5.92 & -1.29 & 0.61 & 2.04 & 0.11 & -1.87 & 1.46 & 1.70 \\
10 & 3412 & \emph{3.79} & \emph{3.79} & 2.50 & 1.75 & 4.96 & -1.14 & -0.12 & -0.46 & -0.09 & 0.06 & -1.98 & -3.74 & 1.22 & 0.62 \\
11 & 6103 & -0.07 & 0.33 & -3.25 & -0.25 & 3.21 & -1.17 & -0.18 & -0.30 & -0.93 & 0.11 & -0.22 & -0.73 & 1.34 & 0.27 \\
12 & 2840 & 0.17 & 1.33 & -0.75 & 1.00 & -3.49 & -5.54 & 0.36 & 0.38 & -2.74 & 1.63 & -1.83 & -1.96 & 0.89 & 1.66 \\
13 & 5760 & 1.37 & 1.87 & 0.25 & -1.00 & 0.83 & -3.00 & -4.73 & -0.23 & 0.08 & 1.99 & -2.26 & -2.43 & 1.29 & 1.03 \\
NS & 2493 & \nodata & \nodata & 1.00 & -0.25 & 1.10 & -0.41 & 2.07 & 0.30 & 1.51 & -0.23 & 1.42 & -0.15 & 0.49 & 0.31 \\
\enddata
\tablecomments{Measurements are in pixels (1 pixel=0\farcs1318).  Data indicate how far the 1999 region is offset from the 2009 
region.  Negative numbers indicate offsets to the east or south.}
\tablenotetext{a}{10-year motions based on optical \citep{bk85} or X-ray \citep[italic,][]{drf04} data.} 
\tablenotetext{b}{Corrected for the proper motions indicated in columns 3 and 4.}
\tablenotetext{c}{Standard deviation of the mean of the four measurements except for the 
neutron star where just the standard deviation is reported.}
\end{deluxetable}

In order to determine the final registration solution between 1999 and 2009, each of the 
$x$- and $y$-offset measures for each region were corrected for their 10-year proper 
motion based on either the optical or X-ray measurement \citep{bk85,drf04}.  The 10-year 
proper motions are indicated in columns 3 and 4 of Table~\ref{regdata}.  About half of 
the QSFs had optical counterparts with very well determined motions such that the 
position error over 10 years is less than 1 HRC pixel.  The other half of the QSFs had 
only X-ray proper motions with larger errors, as much as several pixels over 10 years 
\citep{drf04}.  The proper-motion-corrected offsets were then averaged together for each 
region with the standard deviation of the mean being used as the estimate of uncertainty.

\begin{figure}[t]
\epsscale{1.0}
\plotone{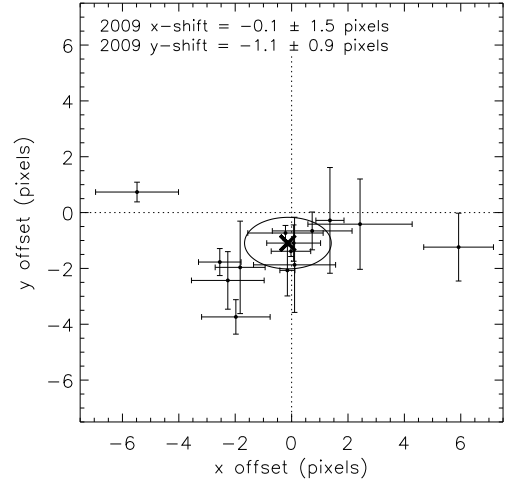}
\caption{The mean offsets and uncertainties for the QSFs and src2 are plotted in HRC 
pixels where 1 HRC pixel = 0\farcs1318.  The final registration solution is indicated by 
the $\times$ and the final registration uncertainty is indicated by the ellipse. 
\label{registration}}
\end{figure}

The weighted mean $x$- and $y$-offsets were calculated using the individual mean offsets 
for each region with the uncertainties used as the weights.  The weighted standard 
deviation is used as the final estimate for the uncertainty in image registration.  The 
final image registration solution is to shift the 2009 image $-0.1\pm1.5$ pixels in $x$ 
and $-1.1\pm0.9$ pixels in $y$ to best match up with the 1999 image.  
Figure~\ref{registration} shows the mean offsets and uncertainties for each of the QSFs 
and src2.  Also shown is the final registration solution, represented by the 
$\times$ and the final uncertainty estimate, represented by the ellipse.

\section{Neutron Star Motion}
\label{motion}

The neutron star position and offsets were measured in the same way as for the QSFs and 
src2 using the same range of binning, smoothing, and bounding box parameters.  The 
individual $x$- and $y$-offsets were averaged together and standard deviations computed.  
These are shown in the last row of Table~\ref{regdata}.  We note that if we simply do not 
apply any registration corrections, the neutron star would appear to move almost due east.

\begin{figure}[t]
\epsscale{1.0}
\plotone{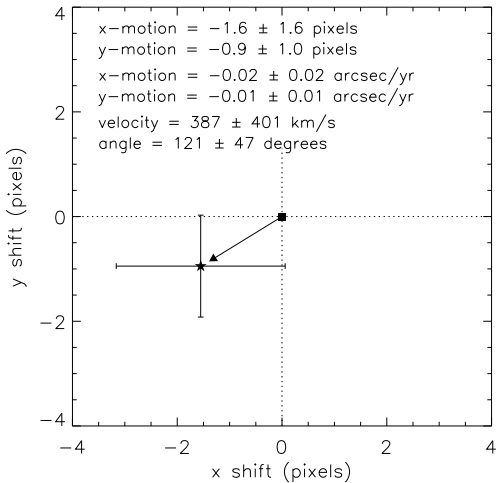}
\caption{The solution for the motion of the neutron star over 10 years. 
\label{nsmotion}}
\end{figure}

\begin{figure}[b]
\epsscale{1.0}
\plotone{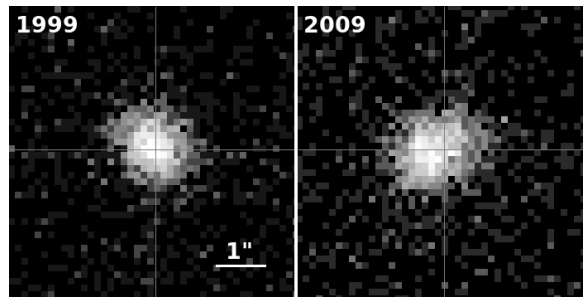}
\caption{Registered 1999 and 2009 images of the neutron star. \label{nscomp}}
\end{figure}

The final solution for the motion of the neutron star was corrected for the registration 
solution and the final uncertainties in the motion are found by adding in quadrature the 
registration uncertainties and the measured offset uncertainties.  This results in a 
proper motion of $-1.6\pm1.6$ pixels ($-0\farcs02\pm0\farcs02$ yr$^{-1}$) in $x$ 
and $-0.9\pm1.0$ pixels ($-0\farcs01\pm0\farcs01$ yr$^{-1}$) in $y$. At a distance of 
3.4 kpc \citep{rhf95}, this results in a velocity of 387$\pm$401 km s$^{-1}$ at 
$121\degr\pm47\degr$ (southeast), which is consistent with the inferred motion of 
350 km s$^{-1}$ at a position angle of $169\degr\pm8.4\degr$ \citep{fes06b}.  
Figure~\ref{nsmotion} shows the final solution and uncertainty for the proper motion of 
the neutron star and Figure~\ref{nscomp} shows final registered images of the neutron 
star in 1999 and 2009.  

\section{Discussion}
\label{disc}

The motion of the neutron star in Cas~A is typical of the birth velocities of young 
pulsars, which range from $\sim200-500$ km s$^{-1}$ \citep{cvb05}.  A number of physical 
mechanisms have been suggested to account for these velocities, including the 
disruption of binaries through mass loss in supernovae \citep{ibe96} and the 
electromagnetic rocket effect \citep{har75}.  The most likely explanation, though, is a 
birth kick imparted to the pulsar through asymmetries in the supernova explosion 
\citep{lai01}.  There is currently no consensus regarding the details of the 
core-collapse explosion process and the resulting asymmetries.  Some models favor 
magnetohydrodynamic instabilities to produce jets which drive a bipolar supernova 
explosion and impart birth kicks parallel to the jet axis \citep{whe02}.  In 
other models, asymmetric neutrino emission, which is mediated by high 
magnetic fields, drives the birth kick \citep{arr99}.

We now know that the neutron star has a low magnetic field \citep{hh09} with a motion 
that is roughly orthogonal to the jet axis.  We know that Cas~A's explosion was 
asymmetric \citep{rfs11} and that there are a number of axisymmetric ejecta alignments 
also roughly orthogonal to the jet axis \citep{err06,srd09,drs10}, including a gap in the 
outer ejecta knot distribution \citep{fes06b}.  Careful mass estimates combined with 
modeling reveal that Cas~A, as a whole, is moving at about 700 km s$^{-1}$ to the north 
\citep{hl12}.  Therefore, hydrodynamic kick mechanisms that do not invoke strong magnetic 
fields or misaligned jets and that only require a modest ejecta mass and velocity 
asymmetry are preferred, such as those of \citet{nbb10, nbb12}.

\section{Conclusion}
\label{conc}

We have attempted to directly measure the proper motion of the neutron star in Cas~A 
over a baseline of 10 years using data from the HRC instrument on \emph{Chandra}.  Due 
to a paucity of background point sources, we were forced to use the QSFs in Cas~A to 
register the 1999 image to the 2009 image.  The QSFs are not point sources, are in 
motion, and changed morphology over the 10-year span.  Thus, our measurement of the 
neutron star motion is marginal at 390$\pm$400 km s$^{-1}$ in the southeast direction.  
This motion agrees with the inferred motion based on the offset of the neutron star from 
the expansion center of Cas~A.  Supernova models that attempt to reproduce the kick 
velocity of the neutron star must also account for the direction of the kick, which is 
roughly orthogonal to the northeast-southwest jet axis in Cas~A.

This work was supported by NASA through Chandra grant GO0-11089X and by the NASA-West 
Virginia Space Grant Consortium.  We thank Shami Chatterjee for his assistance in 
preparing the proposal for the 2009 HRC observations and for reviewing this paper 
prior to submission.

Facilities: \facility{CXO(HRC)}

\end{document}